# Magnetic and charge transport properties of the Na-based Os oxide pyrochlore

Y.G. Shi,[1,*] A.A. Belik,[2] M. Tachibana,[1] M. Tanaka,[3] Y. Katsuya,[4] K. Kobayashi,[3] K. Yamaura,[1] E. Takayama-Muromachi[1,2]

[1] Superconducting Materials Center, National Institute for Materials Science, 1-1 Namiki, Tsukuba, 305-0044 Ibaraki, Japan

[2] International Center for Materials Nanoarchitectonics, National Institute for Materials Science, 1-1 Namiki, Tsukuba, 305-0044 Ibaraki, Japan

[3] NIMS Beamline Station at SPring-8, National Institute for Materials Science, 1-1-1 Kouto, Sayo-cho, Sayo-gun, Hyogo 679-5148, Japan

[4] SPring-8 Service Co. Ltd., 1-1-1 Kouto, Sayo-cho, Sayo-gun, Hyogo 679-5148, Japan

Na-based osmium oxide pyrochlore was synthesized for the first time by an ion-exchange method. $KOs_2O_6$ was used as a host compound. Elelectron probe micro-analysis, synchrotron x-ray diffraction analysis, and thermo-gravimetric analysis confirmed its structure not as the β-type but as the defect-type pyrochlore. The composition was identified as $Na_{1.4}Os_2O_6 \cdot H_2O$. Electrical resistivity, heat capacity, and magnetization measurements of the polycrystalline $Na_{1.4}Os_2O_6 \cdot H_2O$ clarified absence of superconductivity above 2 K, being in contrast to what were found for the β-pyrochlore $AOs_2O_6$ ($A$ = Cs, Rb, K). Sommerfeld coefficient of 22 mJ $K^{-2}$ $mol^{-1}$ of $Na_{1.4}Os_2O_6 \cdot H_2O$ was smallest among those of $AOs_2O_6$. A magnetic anomaly at ~57 K and possible associated magnetoresistance (+3.7 % at 2 K in 70 kOe) were found.

PACS: 74.62.Bf, 74.62.Dh

## I. Introduction

The pyrochlore superconductors $AOs_2O_6$ ($A$ = Cs, Rb, and K) attracted intense attention because of a possible unusual mechanism of superconductivity: rattling of the $A$ atom is believed to play an important role in establishing the superconductivity [1-3]. To date, many studies seem to focus on the possibility; however correlations between the anomalous lattice properties and the superconductivity are still under debate [4-13]. Further additional studies are expected to help to make clear the issue. Besides, it is suggested that $T_c$ goes up efficiently by replacing the $A$ atom by a lighter element. In fact, superconducting critical temperature ($T_c$) goes up step by step depending on the $A$ atom mass from 3.3 K ($A$ = Cs) to 6.4 K (Rb), and to 9.6 K (K) [3], possibly reflecting the unusual mechanism. In order to investigate the mechanism of the superconductivity in $AOs_2O_6$ we tried to synthesize the pyrochlore with Na instated of Cs, Rb, or K. Within our best knowledge, the synthesis was hitherto unachieved.

Regarding strongly correlated electrons properties, Os oxides are fruitful. For example, ferromagnetic Mott transition was observed in $Ba_2NaOsO_6$ [14], metal-insulator transition associated with antiferromagnetic order was found in $Cd_2Os_2O_7$ [15], antiferromagnetic correlations in $Hg_2Os_2O_7$, $Na_3OsO_5$, and $La_2NaOsO_6$ [16-19], and spin-flop transitions in $Pr_2NaOsO_6$ and $Nd_2NaOsO_6$ were notable [19]. Since the observed properties are rather complicated, further development of the Os materials could be useful to shed more light on nature of the materials and thus to establish a comprehensive picture of those. It could also be expected to promote understanding general physics of the $5d$ electrons system.

We carefully tested chemical conditions, aiming synthesis of the Na based pyrochlore. We found that an entire exchange between K and Na was feasible by using the host $KOs_2O_6$. We actually achieved a synthesis of the target and studied its physical properties by measurements of magnetization and charge transport. Here we report primary properties of the Na-based Os oxide pyrochlore for the first time. Although the compound is fairly metallic in nature, an expected superconducting transition was not observed above 2 K. We suggest possible reasons.

## II. Experimental

We prepared the host compound $KOs_2O_6$ by a solid state reaction method from $KO_2$

(powder, Aldrich), $OsO_2$ (Os-84.0%, Alfa Aesar), and Os (99.9%, Aldrich) in a manner reported elsewhere [20]. Powders of $KOs_2O_6$ and $NaNO_3$ (99.9%, Wako) were mixed in a platinum container at 1 to 5 ratio in weight. The mixture was heated in an evacuated quartz ampoule at 330 °C for 7 days. The product was washed in water and dried at 100 °C for 1 hr in air. An electron probe micro-analysis (EPMA) confirmed the composition as a Na rich; 93 % of K ions of the host were replaced by Na. Afterwards, the reaction was repeated in the way, resulting in promotion of the exchange between K and Na. Besides, we observed production of visible $OsO_4$ inside of the ampoule, suggesting an overcharge reaction in the host. The EPMA study was again conducted and revealed that an average metal ratio was Na: K: Os = 1.418: 0.025: 2. The compound is thus labeled as $Na_{1.4}Os_2O_6 \cdot H_2O$ hereafter ($H_2O$ was discussed later). Assuming oxygen composition is stoichiometric, it appears that the Os valence was reduced to +5.3 from +5.5 ($KOs_2O_6$), as a result of accommodation of the excess Na in the host. We did not attempt a thermo-gravimetric analysis (TGA) of the final product above 200 °C, since highly toxic materials $OsO_4$ was possibly released from the sample.

The powder of the polycrystalline $Na_{1.4}Os_2O_6 \cdot H_2O$ was studied by an x-ray diffraction (XRD) method using Cu-$K_\alpha$ radiation in a commercial apparatus (RIGAKU, RINT 2200V). Afterward, detailed crystal structure of $Na_{1.4}Os_2O_6 \cdot H_2O$ was studied by a synchrotron XRD method at room temperature in a large Debye-Scherrer camera at the BL15XU beam line of SPring-8 [21]. Incident beam was monochromatized at $\lambda = 0.65297$ Å, and a sample capillary (Lindenmann glass made, outer diameter was 0.2 mm) was rotated during the x-ray measurement.

An amount of the sample powder was mold into a pellet by a cold press method in a belt-type apparatus (6 GPa was applied without heating). The dense pellet (black in color) was used in charge transport measurements. Electrical resistivity of the pellet was measured by a four-point probe method with ac gage current of 1 to 10 mA at frequency of 30 Hz in a commercial apparatus (Quantum Design, Physical Properties Measurement System). Electrical contacts on the four locations along the pellet were prepared by gold wires and silver paste. Specific heat ($C_p$) was measured by a time-relaxation method in the apparatus. Magnetic properties of the polycrystalline $Na_{1.4}Os_2O_6 \cdot H_2O$ were studied in a commercial magnetometer (Quantum Design, Magnetic Properties Measurement System).

**III. Results and discussion**

Fig. 1 shows the synchrotron XRD pattern collected in a $2\theta$ range between 5° and 60.2° at 0.003° intervals. The obtained pattern was carefully analyzed by a Rietveld method using RIETAN-2000 code [22]. Due to a matter of the XRD background, the lower angle part below 11.4° was unused in the analysis. At the beginning of the analysis, we used the crystallographic parameters of $KOs_2O_6$ (space group: $Fd$-$3m$) as starting parameters [23]. Coefficients for analytical approximation to atomic scattering factors of Os, K, Na, and O were taken from ref. 24. The pseudo-Voigt function of Toraya was used as a profile function [25]. The XRD background was characterized by an 11th-order Legendre polynomial function. Isotropic atomic displacement parameters ($B_{iso}$) and isotropic Debye-Waller factor, $\exp(-B_{iso}\sin^2\theta/\lambda^2)$, were assigned to all the atoms. Because mass of the minor compound $OsO_2$ was too small (estimated ~0.7 % in weight) to measure its crystallographic parameters correctly, we refined only a scale factor and lattice constants in regard to $OsO_2$.

We found significant electron density at not only the $8b$ site but also the $16d$ site in the oxygen cage (see Figs. 2a and 2b), while we started on the assumption that Na ions are only at the $8b$ site. Occupation factors ($g$) and $B_{iso}$ were converged as follows: $g_{Na} = 0.81(4)$ and $B_{iso} = 13.8(9)$ Å$^2$ at the $8b$ site and $g_{Na} = 0.725(9)$ and $B_{iso} = 2.54(12)$ Å$^2$ at the $16d$ site. Since the $B_{iso}$ was unreasonably large, we checked further possible structural models with available chemical species in the cage. As a result, a model with water at the $32e$ site was found better to describe the structure of the compound. The composition deduced from the best solution was $Na_{1.4}Os_2O_6 \cdot H_2O$, which is reminiscent of what was found for $KNbWO_6 \cdot H_2O$ and $KNbWO_6 \cdot 0.69D_2O$ [26]. A TGA result supports the structure model as a notable weight loss more than 3 % was observed when the compound was held in argon at 200 °C overnight, probably due to the dehydration (reaction was not completed). A brief summary of the analysis is in Table I.

Figs. 2a and 2b show, respectively, a schematic view of the crystal structure and an illustration of the two oxygen cages around the nearest $8b$ sites, drawn based on the solution. In the solution, Na ions reside on a $32e$ site in the vicinity of the $16d$ site (see Fig. 2b). The O2 atoms of water molecules also reside on a $32e$ site in the vicinity of the $8b$ site. Although Na$^+$ and $H_2O$ have the same number of electrons, only the combination results in the reasonable solution. If Na is around the 8$b$ site and $H_2O$ is around the 16$d$ site, the bond distance Na-O1 and $H_2O$-O1 becomes unreasonably long and short, respectively. We should note that we refined $g_{Na}$, $B_{iso-Na}$, $g_{O2}$, and $B_{iso-O2}$ independently and simultaneously without any constrain. Besides, the Na content calculated from $g_{Na}$ was in good agreement with the EPMA result. The $OsO_6$ framework was highly rigid with reasonable $B_{iso}$ for Os and O1 sites. The overall refinement converged readily to the fairly reasonable state.

In the superconductor $KOs_2O_6$, K atoms are located exactly at the 8$b$ site, and the 16$d$ site is empty [20]. The specific distribution is so called "β-type". On the other hand, K atoms are located near the 16$d$ site in $KNbWO_6·H_2O$, while $H_2O$ near the 8$b$ site [26]. The hydrated compound is no longer the β-type, but the defect "regular"-type pyrochlore. The "regular" pyrochlore $A_2Tr_2O_7$ has $A$ element at the 16$d$ site and oxygen ($O^{2-}$) at the 8$b$ site. Thus, the compound $Na_{1.4}Os_2O_6·H_2O$ should be classified into rather the defect pyrochlore than the β-pyrochlore. For the Na-based compound $NaNbWO_6·0.5H_2O$, a structural model was proposed with Na at the 16$d$ site and $H_2O$ molecules near the 8$b$ site, being comparable with the present model [26].

A one remarkable feature found in the analysis is that the cubic lattice parameter of $Na_{1.4}Os_2O_6·H_2O$ ($a$ = 10.16958(2) Å) is larger than that of the superconductors $AOs_2O_6$ (10.149 Å for $CsOs_2O_6$ [3], 10.114 Å for $RbOs_2O_6$ [2], and 10.099 Å for $KOs_2O_6$ [1]; see Table II), while Na is much smaller than the other $A$ atoms. Probably, the excess Na accommodation and the hydration account for the result.

Temperature dependence of magnetic susceptibility ($\chi$) of the polycrystalline $Na_{1.4}Os_2O_6·H_2O$ was measured in zero-field cooling (ZFC) and field cooling (FC) conditions between 5 K and 300 K. Applied magnetic field was fixed in a range between 1 kOe and 70 kOe. The $\chi$ vs. $T$ plots are shown in Fig. 3a. An unusual feature was found: the $\chi$ increases step by step with increasing the magnetic field over the whole temperature range, indicating a substantial magnetic background. In order to investigate the magnetic nature, we tried to measure the $\chi$ beyond 300 K. It was, however, unsuccessful because the sample chemically changed.

Subsequently, we measured isothermal magnetization at several temperature points (≤

300 K): the magnetization data are shown in Fig. 3b. The magnetization data clearly show a small ferromagnetic component with a spontaneous magnetization of ~0.001 $\mu_B$ per Os. The component is nearly temperature independent (< 300 K). It is likely that small amount of a ferromagnetic impurity was incorporated in the sample and caused the unusual thermo-magnetic features. We thus carefully repeated the sample preparation to avoid the possibility. Although the XRD analysis did not detect any trace of magnetic impurities, the magnetic background remained visible in all measurements. Alternatively, a parasitic ferromagnetism produced by defects in the host is possibly responsible for the magnetic background, as thoroughly discussed for $Hg_2Os_2O_7$ and $Cd_2Os_2O_7$ [16,17]. A spin canting model proposed for $Na_3OsO_5$ is also possible to account for the background [18]. In any case, further magnetic studies are needed by a microscopic method such as solid state NMR to identify origin of the magnetic background of $Na_{1.4}Os_2O_6 \cdot H_2O$.

Fig. 3c shows a $1/\chi$ vs. $T$ plot of the 10 kOe data. As a Curie-Weiss (CW) feature was found, we applied the CW law to parameterize the linear part (>100 K). It should be kept in mind; however, that the applicability of the CW law to an itinerant system is not theoretically well-justified, although in practice often yields useful results. Fits were conducted with and without a temperature-independent term $\chi_0$. For $\chi_0 = 0$, the best fit yielded 2.84(2) $\mu_B$/Os for the effective Bohr magneton, and -1.17(2)×10$^3$ K for the Weiss temperature, suggesting predominantly antiferromagnetic interactions. For $\chi_0 \neq 0$, the fit yielded 2.90(89) $\mu_B$/Os and -1.20(42)×10$^3$ K at $\chi_0$ = -3.0047(5)×10$^{-5}$ emu/mole. In the plots, an additional magnetic anomaly becomes prominent at 57 K as marked. The anomaly remains visible in the high magnetic field (see Fig. 3a). In the low field ($\leq$ 5 kOe), the magnetic background probably conceals the anomaly. Considering magnetic properties of the related compounds $Cd_2Os_2O_7$ and $Hg_2Os_2O_7$, which show alike features at 227 K [15] and 88 K [17], respectively, it is likely that the anomaly implies establishment of a magnetic order in long range. In order to test the possibility, we carefully measured magneto-charge transport of $Na_{1.4}Os_2O_6 \cdot H_2O$.

Because the ion-exchanged sample was loose powder, we made a dense pellet by a cold press method to make the measurements available. The powder was isotropically compressed at 6 GPa without heating. Fig. 4a shows temperature dependence of the electrical resistivity of the compressed pellet of the polycrystalline $Na_{1.4}Os_2O_6 \cdot H_2O$ measured with and without an applied magnetic field of 70 kOe. Over the whole temperature range between 2 K and 300 K, metallic characters were found such as nearly *T*-linear dependence and low resistivity (~1.2 mOhm·cm at room temperature). The residual resistivity ratio (=$\rho_{300K}/\rho_{2K}$) was moderate 3.2, probably reflecting the polycrystalline nature.

Inset in Fig. 4a shows differential curves of the data. The curves clearly change those characters at ~ 56 K, implying a possible correlation between the charge transport and the magnetic anomaly at 57 K. Besides, the differential curves cut the horizontal zero-axis at ~5 K, indicating a local minimum of the $\rho(T)$ curve. The weak Kondo effect-like feature is probably due to interaction between conducting charges and the magnetic moments.

Over a wide temperature range, manifest *T*-linear dependence was observed (Fig. 4a), suggesting an exotic scattering mechanism [33]. First, we applied the Bloch–Gruneisen (BG) relation to analyze the data, since the model often well describes conventional electron-phonon scattering in a metallic compound [34]. An analytical form was reduced from the BG function to

$$\rho(T) = \rho_0 + \left(\frac{c}{\Theta}\right)\left(\frac{T}{\Theta}\right)^5 \left(120\zeta(5) - \sum_{k=1}^{\infty}\exp(-ky)\left[y^5 + \frac{5}{k}y^4 + \frac{20}{k^2}y^3 + \frac{60}{k^3}y^2 + \frac{120}{k^4}y + \frac{120}{k^5}\right]\right),$$

where $\zeta(p) = \sum_{k=1}^{\infty} k^{-p}$ is the Riemann zeta function, $y = \Theta/T$, $c$ is a constant, and $\Theta$ is the Debye temperature [34]. We used the form to fit the zero-field resistivity data for $Na_{1.4}Os_2O_6 \cdot H_2O$ by a least-squares method; the estimated variable parameters were $\rho_0 = 0.360(2)$ mΩcm, $c = 525(15)$ mΩcm K$^{-1}$, and $\Theta = 211(3)$ K (the best quality fit is shown in Fig. 4b). The deference, $\rho_{exp}-\rho_{cal}$, is shown in the inset of Fig. 4b: a small drop can be seen around 60 K. It is, however, unlikely due to the magnetic anomaly at 57 K because the result $\rho_{exp} < \rho_{cal}$ dose not match any magnetic scattering model. In addition, a comparable order of drop is seen between 200 K and 300 K, suggesting degree of agreement between the data and the model. Throughout the analysis, the model fits the data to some extent, suggesting phonon scattering plays a pivotal role in the charge transport. A possible magnetic scattering at low temperature is too small to be detected in the analysis.

We tested the 70 kOe resistivity data by applying the BG function as well; however a notable progress was unachieved: just a comparable result $\rho_0 = 0.369(2)$ mΩcm, $c = 600(19)$ mΩcm K$^{-1}$, and $\Theta$ = 227(4) K was obtained. The small magnetic field dependence accords with the phonon scattering model. We also tested another practical formula of the BG relation [35], however no significant deference was found.

In Fig 4a, a small degree of positive magnetoresistance (MR) appears near the low temperature limit: We thus decided to investigate the MR features. Fig. 5a shows a magnetic field dependence of the electrical resistivity between 2 K and 70 K: degree of MR increases with increasing the field. Fig. 5b shows a normalized version of the data, indicating saturation of MR at low temperature (approaching to +3.7 % at 70 kOe on cooling). Field dependence of MR was further investigated at several temperature points. The data are shown in the inset of Fig. 5b. The change is monotonic over the field/temperature range studied. Although the MR features are not clearly understood, it is likely that the MR is coupled with the magnetic anomaly at 57 K because the MR emerges just below the characteristic temperature.

We measured specific heat of Na$_{1.4}$Os$_2$O$_6$·H$_2$O to investigate the magnetism further. The $C_p$ data are plotted in a $C_p$ vs. $T$ form in the top panel of Fig. 6. From 2 K to 300 K, $C_p$ changes monotonically without any anomaly. An expected peak was not obvious at the magnetically characteristic temperature 57 K. Although it is likely that undetectably small mass of a magnetic impurity is responsible for the magnetic anomaly, however a magnetic transition without anomaly in heat-capacity could also account for the anomaly, as intensely discussed for Hg$_2$Os$_2$O$_7$ [17]. Besides, unconventional metallic magnetism suggested for LaCrSb$_3$ could also take responsibility for the magnetic feature [36]. In order to test the possible magnetic models, we measured the $C_p$ of Na$_{1.4}$Os$_2$O$_6$·H$_2$O in a magnetic field of 70 kOe. The data are plotted in the main panel of Fig. 6 as well as the zero-field data. It thus becomes clear that there is no remarkable field dependence of $C_p$ over the temperature/field range.

For reference, $C_p$ of the superconductors KOs$_2$O$_6$ and RbOs$_2$O$_6$ are plotted in the same manner in Fig. 6 (the data were taken from ref. 23). The comparison manifests absence of a bell-shaped feature in Na$_{1.4}$Os$_2$O$_6$·H$_2$O. According to ref. 23, the bell-shaped contribution is well characterized by a linear combination of the Debye model and the Einstein model: the Einstein model is mainly responsible for the distinct temperature dependence due to an anharmonic lattice contribution related to the $A$-atom rattling [23]. We attempted to fit the $C_p$ data of Na$_{1.4}$Os$_2$O$_6$·H$_2$O as well, however we found that the Debye model solely accounts for the $C_p$ feature in a reasonable sense without combination to the Einstein model (see the solid curve in the main panel). The solid curve is the best fit obtained by using only 2 variable parameters: the Debye temperature was 228 K and the density was 11.6 modes per the mole. The lack of the Einstein part in the $C_p$ of $Na_{1.4}Os_2O_6 \cdot H_2O$ accords with what was found in the structural analysis: the Na rattling is disabled because it does not crystallize in the β-type.

A low temperature part of the data ($T \ll \Theta$) is plotted in a form $C_p/T$ vs. $T^2$ in the inset of Fig. 6. An expected feature $C_p = \gamma T + \beta T^3$ was clearly observed, in which γ is the Sommerfeld coefficient and β is a coefficient including the Debye temperature. A line fit to the data by a least-squares method (shown as a solid line) yielded $\Theta = 301.9(3)$ K and $\gamma = 21.4(1)$ mJ mol$^{-1}$ K$^{-2}$. The Θ is comparable with those of $AOs_2O_6$ and several related compounds, while the γ is smallest (see Table II). The result suggests that enhancement of the electron effective mass in $Na_{1.4}Os_2O_6 \cdot H_2O$ is not significant.

**IV. Conclusions**

We have successfully achieved an entire ion exchange in $KOs_2O_6$. The Na-based Os oxide pyrochlore was synthesized for the first time and its structure, magnetic and electrical properties were studied. The compound does not turn into a superconducting state above 2 K against an expectation that Na substitution increases $T_c$. Our data suggest several reasons: (1) the compound does not crystallize in the β-pyrochlore structure but the defect "regular"-pyrochlore structure, (2) the Os valence +5.3 is lower than +5.5 of the other superconductors $AOs_2O_6$ (A = Cs, Rb, and K), and (3) the compound is substantially hydrated. Because the compound as made was highly air sensitive and complete dehydration was unachieved, $H_2O$ molecules kept remaining in the cages. Although a very recent work on hydration of the superconductor $KOs_2O_6$ suggested that small degree of the hydration does not kill the superconductivity, the hydrated Na-based one is not a superconductor (> 2 K). Probably because it accommodates $H_2O$ molecules inside 10 times or much denser than the hydrated superconductor $KOs_2O_6 \cdot nH_2O$ ($n < 0.1$) [37].

Considering the structure type, the hydration probably caused a site shift regarding Na from the 8b to the 16d, preventing Na rattling. An analogous shift was found in the hydrated $KOs_2O_6$ [23,26,37]. The hydration likely plays a significant role to destroy the possible superconductivity. Because all the reasons above are caused by the excess Na and the hydration, we investigated more than 10 pieces of preliminary samples prepared in early experimental stages, which may have different quantity of Na and water per the mole. We found, however, none of those showed superconductivity above 2 K.

The compound $Na_{1.4}Os_2O_6 \cdot H_2O$ shows a magnetic anomaly at 57 K without a visible anomaly in heat capacity. The anomaly is coupled with the MR (+3.7 % at 2 K in the presence of 70 kOe). A magnetic transition without a manifest change in magnetic entropy, as discussed for $Hg_2Os_2O_7$ and $LaCrSb_3$ [17,36], probably accounts for the anomaly. Because a possibility of influence of a little magnetic impurity is not completely rejected, further efforts are needed to reveal magnetic nature of the Na-based Os oxide pyrochlore. At the end, we would like to state that it is still possible that a careful control of Na and water quantity in the pyrochlore host produces superconductivity. In order to precisely control the dehydration of the Na-based Os oxide pyrochlore, further studies are in progress.

**Acknowledgments**

We would like to thank Drs. R.W. Li, T. Kolodiazhnyi, and Z. Hiroi for helpful discussions. The research was supported in part by the Superconducting Materials Research Project from MEXT, Japan, the Grants-in-Aid for Scientific Research from JSPS (18655080, 20360012), the Murata Science Foundation, and the Futaba Memorial Foundation.

**References**

* E-mail address: SHI.Youguo@nims.go.jp

Table I. Structure parameters of $Na_{1.4}Os_2O_6 \cdot H_2O$ determined by a synchrotron X-ray powder diffraction method. Space group is *Fd-3m* (no. 227) at origin choice 2, $z = 8$, $a = 10.16958(2)$ Å, $V = 1051.743(3)$ Å$^3$, and $\rho_{cal} = 6.632$ g/cm$^3$. $R$ factors were $R_{wp} = 3.04$ %, $R_p = 1.93$ %, $R_B = 3.53$ %, and $R_F = 1.99$ %. Selected bond lengths are $d_{Os-O1} = 1.927(1)$ Å (×6), $d_{Na1-O1} = 2.556(3)$ Å (×3), $d_{Na1-O1} = 2.611(4)$ Å (×3), and $d_{Na1-O2} = 2.256(15)$ Å.

| Site | Wyckoff position | g | x | y | z | B (Å$^2$) |
|---|---|---|---|---|---|---|
| Os | 16*c* | 1 | 0 | 0 | 0 | 0.170(5) |
| Na1 | 32*e* | 0.356(4) | 0.4900(9) | = x | = x | 1.56(19) |
| O1 | 48*f* | 1 | 0.3183(3) | 0.125 | 0.125 | 0.57(7) |
| O2 (H$_2$O) | 32*e* | 0.248(10) | 0.4048(8) | = x | = x | 1.1(5) |

Table II. Superconducting transition temperatures ($T_c$), lattice parameters ($a$), the Sommerfeld coefficients ($\gamma$), and the Debye temperatures ($\Theta$) of the pyrochlore Os oxides (above the separator) and related pyrochlore oxides.

| Material | $T_c$ (K) | $a$ (Å) | $\gamma$ (mJ mol$^{-1}$ K$^{-2}$) | $\Theta$ (K) | reference |
|---|---|---|---|---|---|
| Na$_{1.4}$Os$_2$O$_6$·H$_2$O | ---- | 10.16958(2) | 21.4(1) | 301.9(3) | This work |
| KOs$_2$O$_6$ | 9.5 | 10.0968(8) | 76-110 | | [23] |
| RbOs$_2$O$_6$ | 6.3 | 10.1137(1) | 40.4-44 | 325 | [23,27,28] |
| CsOs$_2$O$_6$ | 3.3 | 10.1477(1) | 39.0-40.0 | | [27] |
| Cd$_2$Os$_2$O$_7$ | ---- | 10.1651(4) | 1.1-1.4 | 354-463 | [15] |
| Hg$_2$Os$_2$O$_7$ | ---- | 10.228 | 21 | 230 | [17,29] |
| Cd$_2$Re$_2$O$_7$ | 1 | 10.226(2) | 26.5-30.2 | 285-458 | [30,31] |
| Cd$_2$Ru$_2$O$_7$ | ---- | 10.129(1) | 12.3 | 310 | [31,32] |

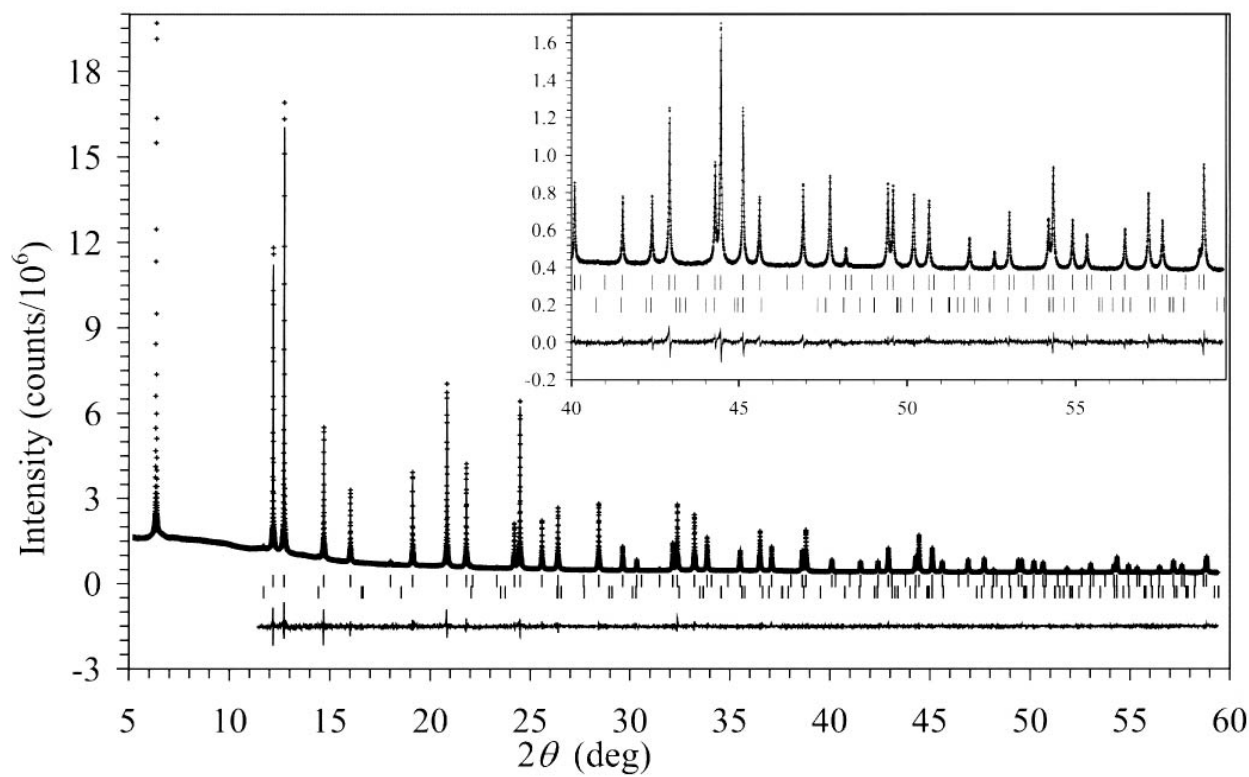

Fig. 1　Result of Rietveld analysis of $Na_{1.4}Os_2O_6 \cdot H_2O$ with the powder diffraction data obtained by synchrotron X-rays. Cross markers and solid lines show the observed and calculated powder diffraction profiles, respectively, and the difference is shown at the bottom.　The positions of Bragg reflections are marked by ticks.　The lower line of ticks is for the impurity $OsO_2$.　Inset shows an expansion of the pattern.

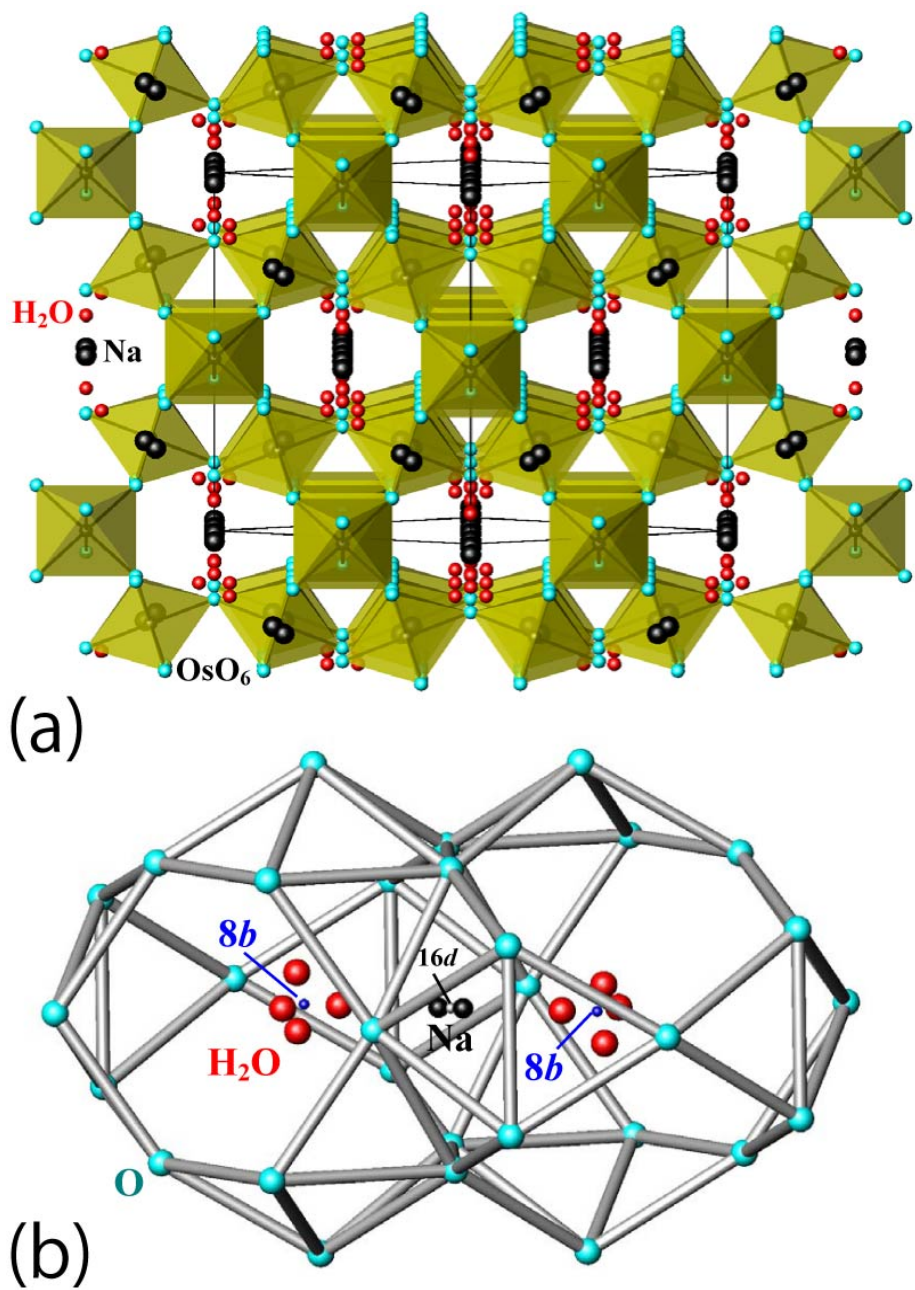

Fig. 2 (a) View of the crystal structure of $Na_{1.4}Os_2O_6 \cdot H_2O$ along the [110] direction. $OsO_6$ octahedra are shown. Large circles in channel are Na1 atom, and small circles are O2 ($H_2O$) atom. (b) Two oxygen cages are combined around the 16d site. Na1 atoms and O2 ($H_2O$) atoms are at the 32e site in the vicinity of the 16d site and the 8b site, respectively.

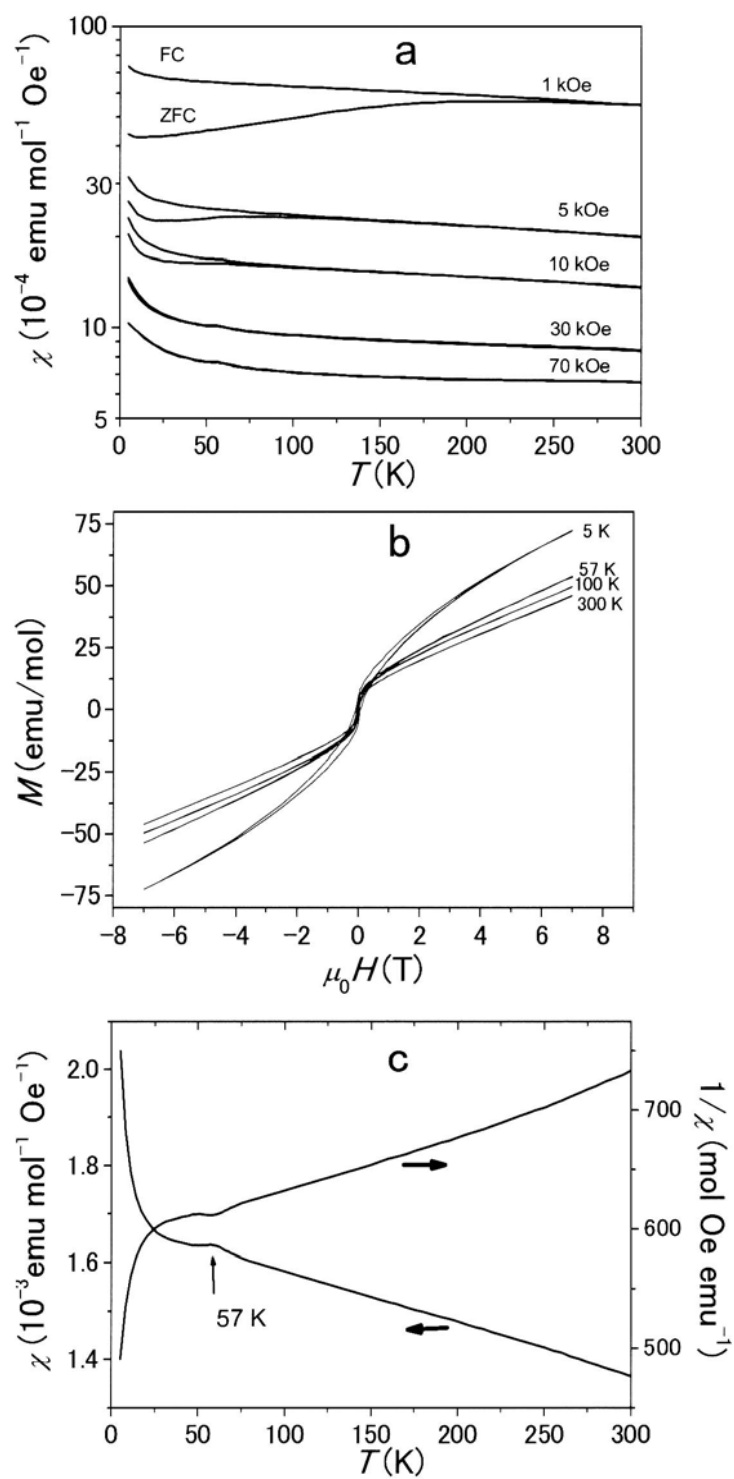

Fig. 3 (a) Temperature dependence of the magnetic susceptibility of the polycrystalline Na$_{1.4}$Os$_2$O$_6$·H$_2$O, (b) isothermal magnetization at various temperatures, and (c) an alternative plot of the susceptibility data measured at $H$ = 10 kOe.

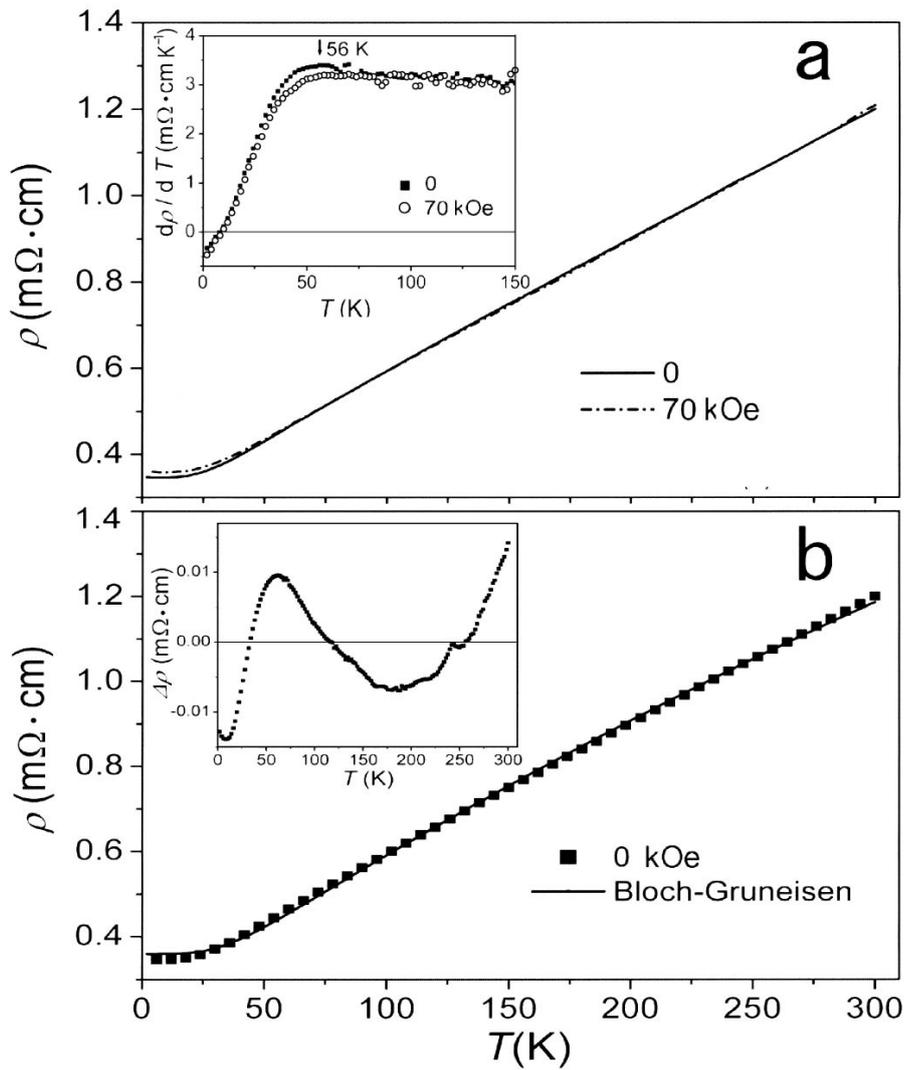

Fig. 4 (a) Temperature dependence of electrical resistivity of the polycrystalline $Na_{1.4}Os_2O_6 \cdot H_2O$ measured at $H = 0$ and 70 kOe.  Top inset shows differential curves of the data. (b) The Bloch–Gruneisen function (solid curve) and the $H = 0$ data points.  Bottom inset shows difference between those.

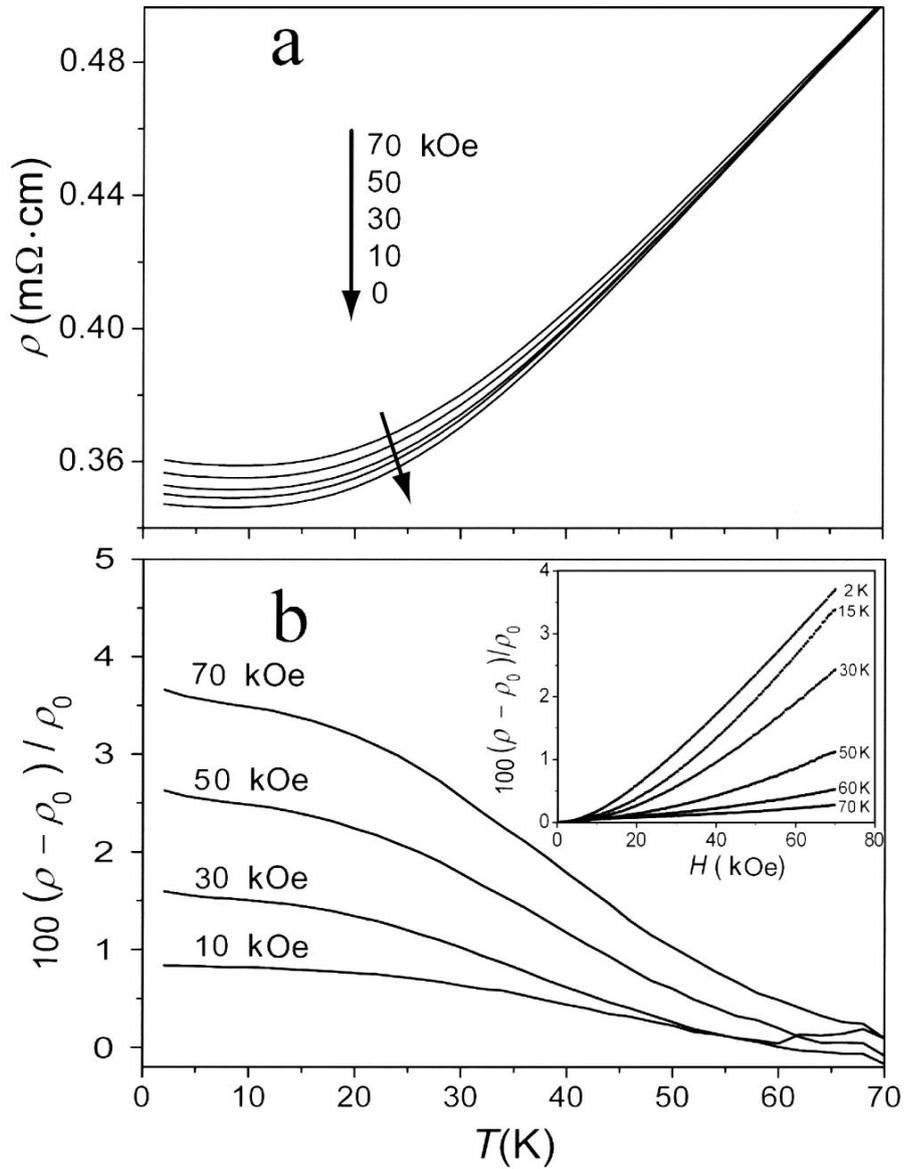

Fig. 5 (a) Field dependence of electrical resistivity of the polycrystalline $Na_{1.4}Os_2O_6 \cdot H_2O$ at low temperature, and (b) the normalized plot. Inset shows the $\Delta\rho/\rho(0) - H$ relation at various temperatures. Note that the $\Delta\rho/\rho(0)$ below ~70 K are positive.

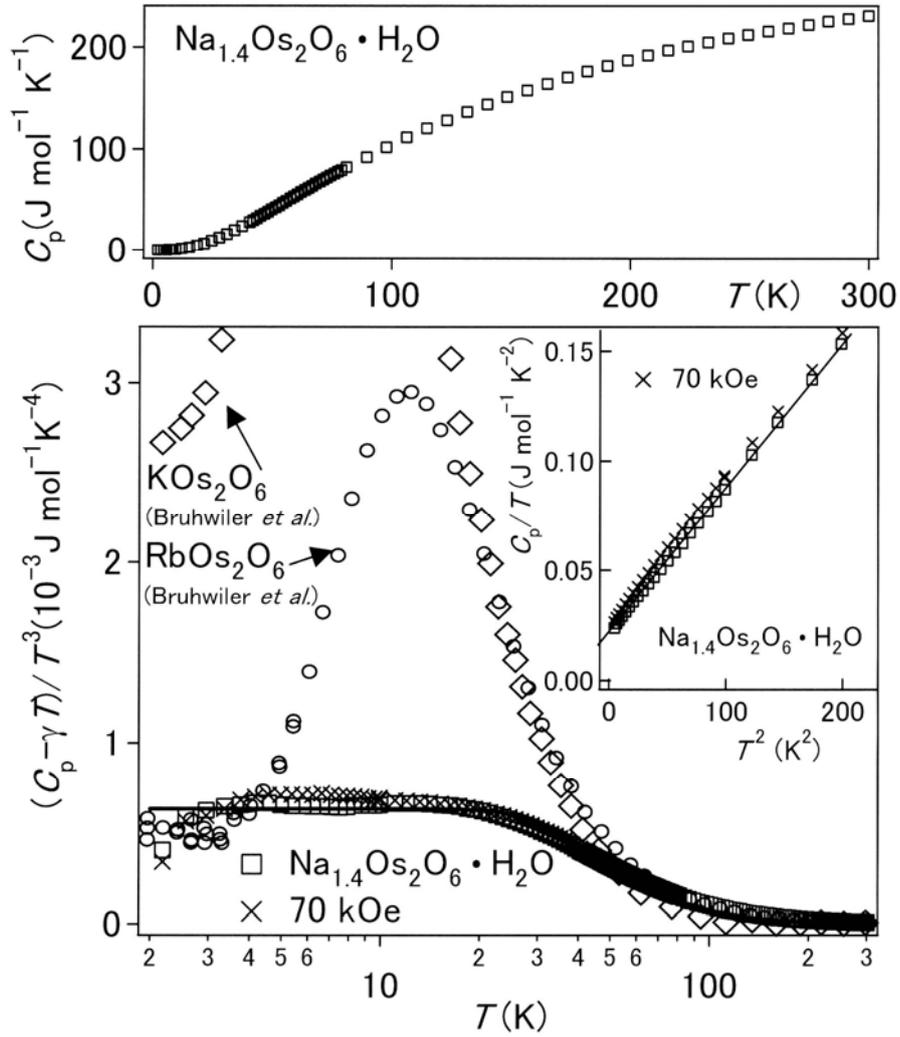

Fig. 6  Heat capacity of $Na_{1.4}Os_2O_6 \cdot H_2O$ at $H = 0$ kOe and 70 kOe.  The upper panel shows $C_p$ vs. $T$ plot between 2 K and 300 K.  The main panel shows a comparison of the lattice heat capacity $(C_p - \gamma T)/T^3$ with those of $KOs_2O_6$ and $RbOs_2O_6$ (the data were taken from ref. 23).  The inset shows $C_p/T$ vs. $T^2$ plots of the $Na_{1.4}Os_2O_6 \cdot H_2O$ data.